\documentclass[onecolumn,secnumarabic,amssymb,  aps, prl, reprint]{revtex4-1}

\usepackage{amsmath}
\usepackage{graphicx}
\usepackage{bm}
\usepackage[usenames,dvipsnames]{color} 
\usepackage{epstopdf}

\begin{document}

\title{Direct measurements reveal non-Markovian fluctuations of DNA threading through a solid-state nanopore}

\author{Nicholas A. W. Bell$^{1}$}
\email[correspondence: ]{nawb2@cam.ac.uk}
\author{Ulrich F. Keyser$^{1}$}
\email[correspondence: ]{ufk20@cam.ac.uk}
\affiliation{$^{1}$Cavendish Laboratory, University of Cambridge, CB3 0HE, UK}

\begin{abstract}
The threading of a polymer chain through a small pore is a classic problem in polymer dynamics and underlies nanopore sensing technology.  However important experimental aspects of the polymer motion in a solid-state nanopore, such as an accurate measurement of the velocity variation during translocation, have remained elusive. In this work we analysed the translocation through conical quartz nanopores of a 7~kbp DNA double-strand labelled with six markers equally spaced along its contour.  These markers, constructed from DNA hairpins, give direct experimental access to the translocation dynamics.  On average we measure a 5\% reduction in velocity during the translocation. We also find a striking correlation in velocity fluctuations with a decay constant of 100s of $\mu$s.  These results shed light on hitherto unresolved problems in the dynamics of DNA translocation and provide guidance for experiments seeking to determine positional information along a DNA strand.
\end{abstract}

\maketitle

\begin{center}
 \textbf{I. Introduction}
\end{center}
Ionic current sensing with nanopores is a powerful technique for probing the structure and dynamics of charged polymers in confined geometries.  Solid-state nanopores can be fabricated with a range of geometries which enables them to characterise a wide spectrum of biopolymers such as DNA, RNA and proteins.  A substantial effort has been made in the field towards the goal of using solid-state nanopores for DNA sequencing \cite{Branton2008,Wanunu2012}.  There are also other potential applications which rely on detecting the positions of bound objects along the contour of DNA such as sequence specific probes \cite{Singer2012} and proteins \cite{Kowalczyk2010b, Shim2015,Yu2015,Bell2016a}.   These developing applications, as well as future sequencing methods, require a quantative understanding of the average spatial trajectory and fluctuations of DNA in order to determine positional accuracy.

Most experimental and theoretical efforts have attempted to understand DNA translocation by determining the scaling  of the average translocation time $\bar{t}$ with DNA length $N$ according to $\bar{t} \sim N^{\alpha}$.  In nearly all experimental reports for solid-state nanopores, double-stranded (ds)DNA several thousand basepairs in length is used for which the Zimm relaxation time is significantly longer than the translocation time.   In this regime the threading must be a non-equilibrium process. Experimental results (with relatively wide 8~nm-15~nm diameter nanopores) have measured a spread of values for the scaling exponent $\alpha$ typically between 1.2 and 1.4 \cite{Storm2005a, Li2010, Mihovilovic2013} although some results indicate a linear scaling \cite{Chen2004}. Theoretical models and simulations for non-equilibrium translocation have given predictions of a range of $\alpha$ which are greater than 1 \cite{Panja, Palyulin2014a}. The differences in $\alpha$ may partly reflect variations in conditions such as the relative contribution of friction inside the pore compared to that outside \cite{Ikonen2012a}.  In general however the ongoing difficulty in comparing experimental scaling laws with theory motivates more direct ways of characterising DNA translocation dynamics.

Several studies have analysed another important aspect of translocation dynamics namely the width of the distribution of translocation times for a single DNA length.  These studies have found that  the spread in translocation times  is underestimated by simple models of biased diffusion \cite{Lu2011}.  Furthermore biased diffusion can not account for the dependence of the distribution width with voltage \cite{Chen2004, Li2010}. In a recent experiment \cite{Plesa2014}, DNA constructs with a long protrusion at a known position were measured and exhibited surprisingly large translocation fluctuations.

In this paper we use DNA dumbbell motifs to create multiple markers along a DNA double-strand and thereby accurately determine the  dynamics during threading through a nanopore.  We designed a DNA double-strand which has six equally separated zones of dumbbells creating positional markers which are read with high signal to noise.  The translocation times of these markers were measured with statistics of thousands of translocations at multiple voltages.  During the translocation, excluding start and end effects, we are able to measure a 5$\%$ reduction in velocity.  Furthermore we demonstrate that there is a correlation in the DNA velocity on a timescale comparable to the total DNA chain translocation time and hence that the DNA fluctuations are non-Markovian. We discuss potential mechanisms for the observed dynamics and their implications for experiments seeking to determine position along a DNA double-strand.

\begin{center}
 \textbf{II. Methods}
\end{center}

\textbf{DNA synthesis}

Our design aim was to fabricate a dsDNA molecule with multiple markers along its contour at known positions. These markers should cause minimal perturbation to the translocation but still be sufficient for high signal to noise reading.  Previous analysis showed that units of DNA dumbbell hairpins \cite{Bell2016a} can be used to create such markers.  For this study, we created a design whereby six zones of DNA dumbbells were placed along a 7228~bp backbone (Figure 1a).  Each zone contained eight dumbbells with each dumbbell made by 28~bases protruding from the dsDNA backbone (Figure 1a).  Therefore each zone has a total molecular weight of protrusions of 148~kDa which is significantly less than the overall 4.8~MDa weight of the 7228~bp DNA double-strand and we consider that this will not cause a significant change to the dynamics of a bare double-strand.  Furthermore the dumbbells wrap around the DNA double helix so that the construct is not wider than the diameter of the nanopores used here.  The design is symmetric so that translocations in either polarity can be considered as equivalent.

We note that nicks (ie single breaks in the DNA phosphate backbone of one strand) occur at 38~bp intervals along the majority of the DNA (and at 10~bp separation in the dumbbell zones).  Numerous studies have investigated the effect of nicks on dsDNA structure and dynamics and consistently observed no significant difference between non-nicked and nicked DNA under high salt concentrations as used in this study \cite{Snowden-Ifft1990,Mills1994,Furrer1997}.  Indeed we previously measured indistinguishable nanopore translocation dynamics between a 7.2~kbp DNA construct with 38~bp interval nicks and a purified 7~kbp DNA plasmid \cite{Bell2015}.  Therefore we consider that our dsDNA construct behaves similarly to non-nicked dsDNA.     

\textbf{Nanopore measurements}

Nanopores were fabricated by laser assisted pulling of glass capillaries with an established protocol which yields final tip diameters estimated by electron microscopy as 14$\pm$3~nm (mean$\pm$s.d.) \cite{Bell2016a}. The nanopore forms a conical geometry - the average geometry is shown in Figure 1b with an example image shown in 1c.  Each nanopore was filled with 10~mM Tris-HCl (pH=8), 1~mM MgCl2, 50~mM NaCl and 4~M LiCl.  DNA was added at a final concentration of 2-6~nM.  Translocations were recorded using an Axopatch 200B and an external 8-pole Bessel filter set to 50~kHz with a sampling rate of 250~kHz.  Both reservoirs were sealed during the experiment and we only used experiments where the baseline drift was $<$0.5$\%$ for the entire recording. This means we exclude experiments where there are significant baseline shifts or `clogging' which can affect the translocation properties \cite{Liang2015}.  

Thousands of translocations were recorded across multiple nanopores and at a variety of voltages.  We then employ several stages of selection to remove unwanted translocation events which are due to folded DNA configurations and fragments of the full length.  Firstly we perform a threshold analysis to select only events that pass 100~pA from the baseline.  Secondly, we plot a histogram of the event charge deficit (ECD) and only select events  3$\sigma$ either side of the Gaussian peak in ECD.  ECD gives a measure of the molecule length \cite{Bell2016} and this step removes the $\sim$20$\%$ of translocations that are due to fragments of DNA (a consequence of the synthesis procedure which was extensively characterised previously \cite{Bell2015}). Thirdly, we remove events where the DNA has a folded configuration by requiring that the DNA does not cross a defined threshold within the first 5$\%$ of the total translocation time.  This removes so called type `21' folds which are the predominant non-single file threading conformation \cite{Mihovilovic2013,Storm2005}.  Finally we use an automated peak fitting algorithm to determine the positions of the intra-event peaks due to the markers.  In this final stage 81-92$\%$ of translocations are identified as containing six peaks with the remainder being discarded. We note that we do not use the start and end of the translocation as time markers.  We do this primarily since the start and end signals are not as well defined as the peaks due to the markers thus preventing determination of their timings with high precision. Also it is possible that there remain some fragments (not excluded by the ECD filter step) which could influence the measurement for the start and end.

\begin{figure}
 \includegraphics{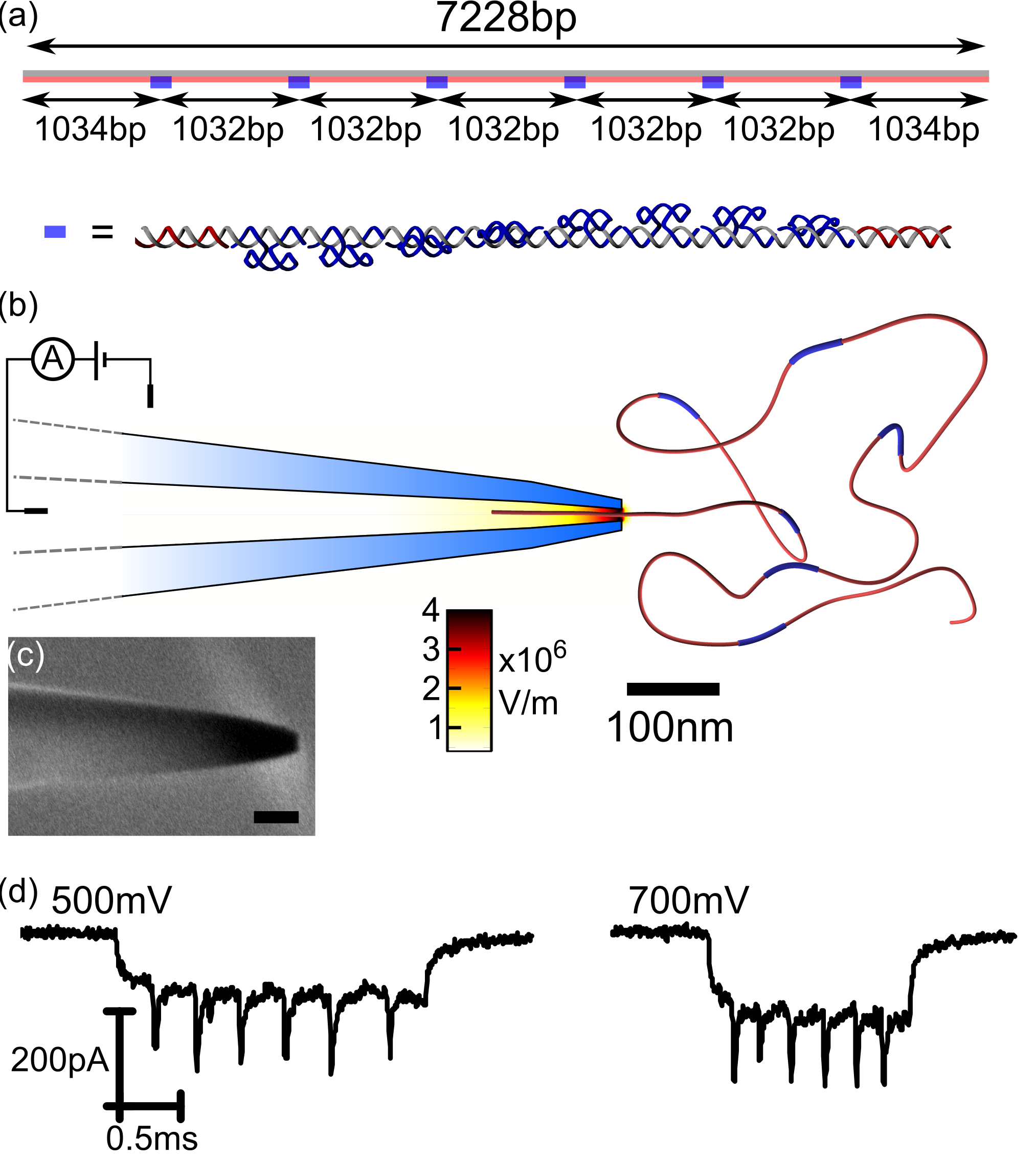}
 \caption{\label{Figure1} (a) Outline of DNA construct showing positions of DNA dumbbell hairpin zones along the backbone and 3D structure of zones. (b) Schematic of conically shaped nanopore (to scale) with average geometry determined by scanning electron microscopy (SEM).  The electric field strength was calculated using Poisson-Nernst-Planck equations in a 2D axisymmetric geometry with COMSOL multiphysics.  (c) Example SEM image showing the outer dimensions of the conical nanopore. Scale bar = 100~nm. (d) Example ionic current recordings of DNA translocations at 500~mV and 700~mV.}
 \end{figure}

\begin{center}
 \textbf{III. Results}
\end{center}

Firstly we consider the average passing time of the markers.  Figure 2a shows our definitions of intra-marker times $\tau_i$.  In Figure 2b the distribution of two of these times is plotted from a total of 872 translocations with a single nanopore at 700~mV (Pore6 in Figure 2c).  The mean time of the $\tau_i$ distribution is given as a function of $i$ for six nanopores in Figure 2c.   The velocity varies for different nanopores which can be expected from small variations in geometry and therefore electric field profile within the nanopore.  We observe a small but consistent increase in $\bar{\tau_i}$ with $i$ - there is an average increase of 5$\%$ between $\bar{\tau_5}$ and $\bar{\tau_1}$.  Since the position intervals between the markers is constant by the DNA design this shows that the velocity tends to decrease during the translocation.

\begin{figure}
 \includegraphics{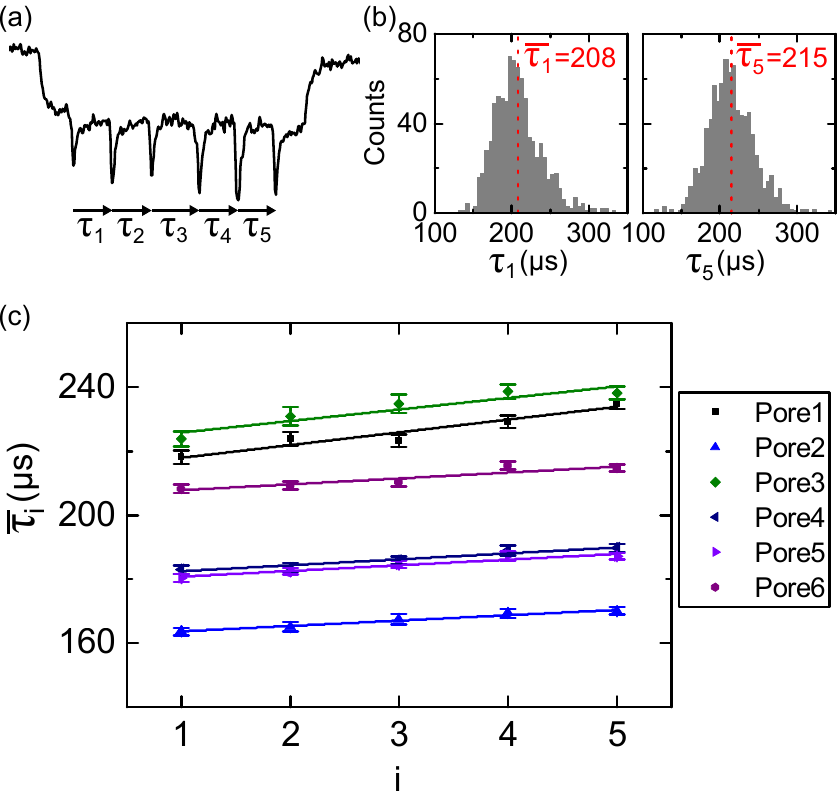}
 \caption{\label{Figure2} Determination of average trajectory of a DNA molecule. (a) Example translocation showing definition of intra-event intervals $\tau_1$ to $\tau_5$.  (b) Histograms of $\tau_1$ and $\tau_5$ for one particular nanopore at 700~mV applied potential (N=872). Each count is an individual translocation. (c) Mean time as a function of i for multiple nanopores at 700~mV where i is the subscript as defined in (a). Error bars represent standard error of the mean. A least-squares linear fit is shown for each data set.}
 \end{figure}

We also analysed the data for correlations in the intra-event dynamics. Figure 3a shows scatter plots of $\tau_5$ versus $\tau_4$ and $\tau _5$ versus $\tau _1$. We quantify the correlation using the Pearson correlation coefficient given by $\rho(\tau_{i},\tau_{j})=cov(\tau_{i},\tau_{j})/\sigma_{\tau_{i}} \sigma_{\tau_{j}}$.  A clearly non-zero correlation is observed for $\tau_5$ versus $\tau _4$ with a lower correlation seen for $\tau _5$ against $\tau _1$ where the intervals are further apart along the DNA chain.   There are ten total possible combinations of $\tau_i$ and $\tau_j$ and in Figure 3b we plot the value of the correlation coefficient against the average time between the intervals given by $\sum_{n=1}^{j} \bar{\tau}_n  - \sum_{m=1}^{i} \bar{\tau}_m$.  Data from four voltages is combined yielding a total of forty data points.  We observe a  decrease in correlation coefficient as a function of the average time between intervals with a consistent behaviour measured across all voltages. A phenomological exponential decay of the form  $y=A e^{-t/B}$ is fitted by the least-squares method and gives a decay constant B=352 $\mu s$.  The correlation is therefore significant when compared with the total translocation time of the DNA especially at higher voltages.  For example at 700~mV the total translocation time is $\sim$1.5~ms (Figure 1d).

\begin{figure}
 \includegraphics{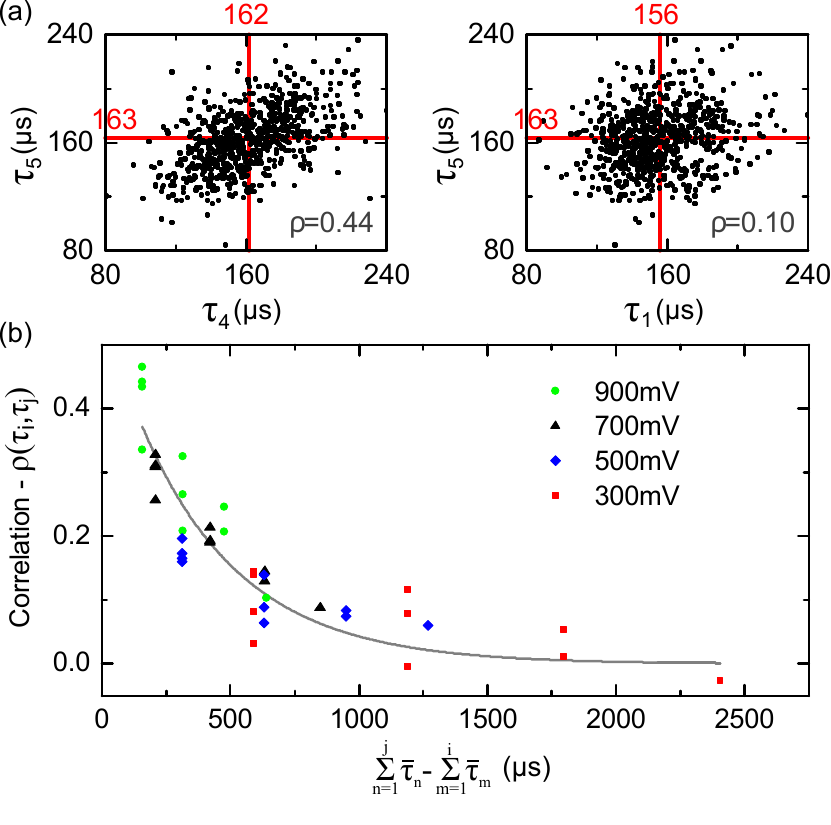}
\caption{\label{Figure3} Intra-event motion correlation.  (a) Scatter plots comparing $\tau_i$ values (as defined in Figure 2a) for Pore6 at 900~mV applied potential with N=727 total data points. (b) Correlation coefficient as a function of average time separation between intervals for Pore6. $^{5}C_2$ = 10 combinations of $\tau_i$ and $\tau_j$ are possible at each voltage.  The total number of translocations are N=727 (900~mV), N=872 (700~mV), N=1020 (500~mV) and N=644 (300~mV).}
\end{figure}

We also determined the cumulative spread in marker position measured with respect to the first marker (Figure 4).  This is directly pertinent for determining the accuracy with which objects bound to a DNA can be located.  The distribution of times shows a slight positive skew as can be seen in Figure 4b. We calculated the variance of each distribution and plotted this as a function of the average translocation time (Figure 4c).  A power law scaling of $\sigma ^2 (t) \sim \bar{t}^\gamma$ was calculated for six nanopores yielding an average value of $\gamma$=1.5. 

The observed scaling for the variance of the translocation time can be rationalised based on our previous observation of a long-lived velocity correlation. In the quasi-equilibrium model of polymer translocation over an entropic barrier, the variance of the translocation time is proportional to the mean translocation time, $\sigma ^2 (t) \sim \bar{t}$ \cite{Muthubook}.  This is the same scaling as the mean first passage time of a particle undergoing Brownian motion with a strong background drift.  The fluctuations of both these models are  Markovian so that $\langle \rho(v(t)v(0)) \rangle = \delta(t)$.  Conversely if the regime is completely correlated motion the translocation time would be set by the initial starting velocity of the DNA and therefore the variance would scale as $\sigma ^2 (t) \sim \bar{t}^2$.  For a finite decaying correlation with a time constant on the order of the translocation time, as observed in these experiments, we would expect a power law scaling in between these two limits which is indeed what is observed.
We note that Chen \textit{et al.} \cite{Chen2004} measured the dependence of $\sigma ^2 (t) \sim \bar{t}^\gamma$ with similar diameter nanopores by recording the translocation time of 48~kbp DNA, at multiple voltages, in 1~M KCl where the translocation speed is approximately ten times faster than 4~M LiCl used in this paper \cite{Bell2016,Kowalczyk2012a}.  They found a scaling $\sigma ^2 (t) \sim \bar{t}^2$ which suggests that with faster translocation speed the motion becomes even more strongly correlated.

\begin{figure}
 \includegraphics{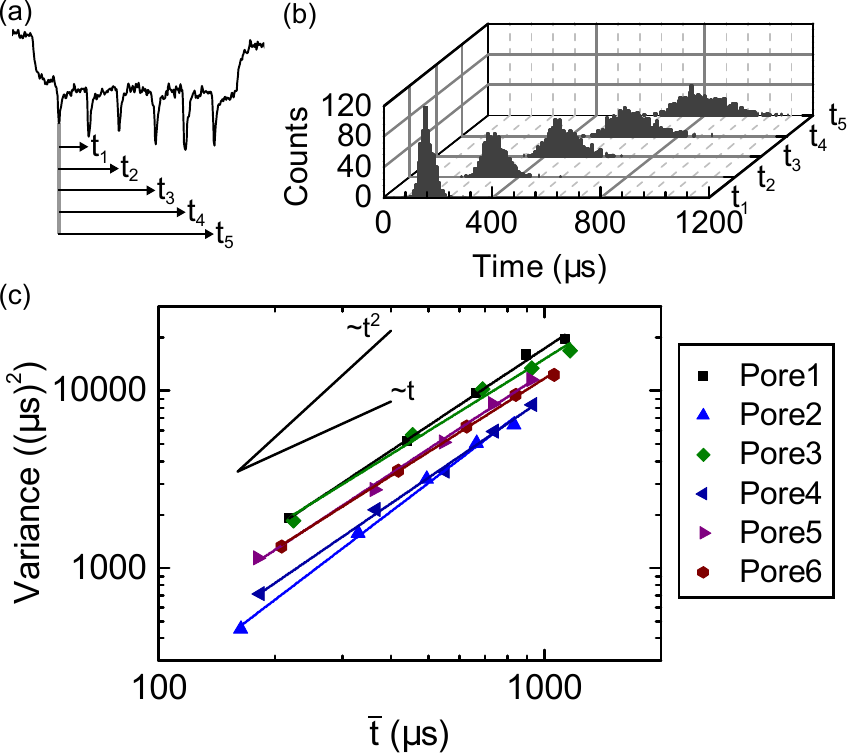}
 \caption{\label{Figure4} Cumulative spread in translocation time. (a) Definition of cumulative times $t_1$ to $t_5$ measured with respect to the first marker to pass through the nanopore.  (b) Histograms showing distributions of times defined in (a) using the same nanopore data as Figure 3 at 700~mV (N=872).  (c) Variance against average time for six nanopores all recorded at 700~mV. There are five data points for each nanopore corresponding to the five times defined in (a).  The average scaling, independently fitting each nanopore data set, is $\bar{t}^{1.5}$. The slopes of quadratic and linear scaling are plotted as a guide to the eye.}
 \end{figure}

\begin{center}
 \textbf{IV. DISCUSSION}
\end{center}
We have demonstated two important features of DNA translocation namely a slight reduction in average velocity during translocation and a long-lived velocity correlation.  A key point to consider for interpreting these results is how the transport time compares with the Zimm relaxation time of the DNA molecule given by \cite{Doi1996}
\begin{equation}
\tau_{Zimm}=\frac{0.3 \eta (\sqrt{N} l_0)^3}{k_B T},
\end{equation}
where $\eta$ is the viscosity, $N$ is the number of Kuhn segments, $l_0$ is the Kuhn length.    For dsDNA, as considered in this experiment, the Kuhn length decreases with increasing salt concentration and tends to a limiting value of approximately 60-100 nm \cite{Sobel1991,Savelyev2012}.  Using the lower limit of 60~nm, together with $\eta=1.7$x$10^{-3}$ ~Pas for 4M LiCl \cite{Tanaka1991} and a DNA length of 7.2 kbp yields a value of $\tau_{Zimm}= 7$~ms which is  longer than the translocation times measured at all voltages. This shows that longer modes of the DNA will be effectively frozen in place and the translocation must be considered as a non-equilibrium process \cite{Palyulin2014a,Panja, Muthubook}.  

The average DNA velocity, $v(t)$, is then given by force balance; $v(t)=F/\gamma(t)$ where $F$ is the electrophoretic force and $\gamma(t)$ is the friction coefficient.  We assume $F$ is constant since the high electric field strength is confined to a few 100~nm region at the nanopore tip (Figure 1b) and in our experiments we do not include the first and last 1034~bp (352~nm) of the translocation.  $\gamma(t)$ can then be decomposed into two parts; $\gamma(t) = \gamma_{pore} + \Gamma(t)$ where $\gamma_{pore}$ is the time independent friction inside the few 100~nm high electric field strength zone and $\Gamma(t)$ is the friction associated with movement of DNA outside this region.   A hydrodynamic model of $\gamma_{pore}$ predicts a velocity that is comparable to that observed in experiments \cite{Ghosal2006} showing that this can account for a substantial proportion of $\gamma(t)$. 

The observation of a slightly increasing $\Gamma(t)$ can be explained by two mechanisms.  Firstly as the translocation progresses an increasing length of DNA is straightened out into the conical geometric confinement thereby increasing drag.  This portion of the DNA on the `trans' side of the translocation is usually neglected in models of polymer transport through thin 2D membranes due to the assumption that it buckles under compression.  However in the conical nanopore geometry used here the DNA is strongly confined and therefore may continue to exert a degree of drag even when it is no longer under significant tension.  The second potential mechanism of increasing $\Gamma(t)$ is associated with DNA segments outside the pore in the `cis' region.  Here as the translocation progresses successive loops of DNA are set into motion as tension propagates through the DNA leading to an increase in the hydrodynamic drag over time \cite{Grosberg2006,Palyulin2014a,Sakaue2010}. The relative contributions of the two mechanisms described above remains to be determined but future comparison with nanopores in 2D membranes should help with assessing the importance of the `trans' confinement in the conical geometry. 

We now consider the nature of the correlated fluctuations shown in Figure 3.  Lu \textit{et al.} \cite{Lu2011} experimentally identified a source of non-thermal fluctuations by showing that the distribution of translocation times for a single DNA length is wider than expected based on a fluctuation-dissipation theorem calculation.  They proposed that additional  fluctuations are caused by the random initial configurations of the DNA at the start of translocation and the associated differences in viscous drag.  Recent theoretical calculations and simulations have also shown the importance of initial configurations in creating additional fluctuations in non-equilibrium, driven DNA translocation \cite{Sarabadani2014,Saito2012,Dubbeldam2013}. 
Our intra-molecule position markers allow us to directly assess the nature of DNA translocation fluctuations and observe both super-diffusive behaviour and a long-lived velocity correlation.  This provides clear confirmation of the presence of non-thermal fluctuations.  
Coarse-grained molecular dynamics, taking into account configurational stochasticity, have made predictions of both these aspects viz. super-diffusive motion and a long-lived velocity correlation \cite{Dubbeldam2013}.  Our study thereby lends support to the importance of configurational stochasticity as an important fluctuation source.

We note that in these coarse-grained molecular dynamics simulations and in our discussion so far we have neglected any potential fluctuations from attractive surface interactions between the DNA and nanopore \cite{Luan2015}.  Such interactions are known experimentally to have an influence for small diameter $\lesssim$ 5nm nanopores where the distribution of translocation times is signifantly wider than for $\gtrsim$ 8nm diameter nanopores \cite{Wanunu2008}.  The nanopores used here are relatively wide at 14$\pm$3~nm and show narrow dwell time distributions \cite{Bell2016} which indicates that we are in the regime where complex surface interactions are minimal.  It will of course be interesting to use our designed DNA rulers to probe the dynamics in the small diameter regime where surface interactions likely dominate.

\begin{center}
 \textbf{V. CONCLUSIONS}
 \end{center}

To conclude, we have shown a new method of designing DNA molecules with multiple position markers for accurately determining the average velocity and fluctuations during DNA translocation through a nanopore. We measure a $\sim$5$\%$ decrease in average velocity during the translocation likely due to an increasing hydrodynamic drag as the translocation progresses. Furthermore we have shown for the first time that a strong velocity correlation exists for adjacent segments of the chain.  This non-Markovian behaviour gives rise to a super-diffusive scaling for the translocation time variance along the DNA chain. These results give new insight into the physics of DNA translocation and enable quantatitive predictions for positional accuracy in nanopore experiments.  

\begin{center}
 \textbf{ACKNOWLEDGEMENTS}
\end{center}
\begin{acknowledgments}
We thank  K. Misiunas, S. Ghosal, K. Chen, M. Muthukumar, and J. Cama for useful discussions and acknowledge funding from an ERC consolidator grant.
\end{acknowledgments}

\bibliographystyle{apsrev}
\bibliography{Library}

\end{document}